%% file: main.tex
\documentclass[conference,compsoc]{IEEEtran}
\IEEEoverridecommandlockouts
\author{
Alireza Lotfi$^{*}$,
Mirza Masfiqur Rahman$^{*}$,
Imtiaz Karim$^{\dagger}$,
Elisa Bertino$^{*}$
\\[0.5ex]
$^{*}$\textit{Purdue University},
$^{\dagger}$\textit{The University of Texas at Dallas}
\\
$^{*}$\{\textit{lotfia, rahman75, bertino}\}@purdue.edu,
$^{\dagger}$\textit{imtiaz.karim}@utdallas.edu
}

\usepackage{graphicx}
\usepackage{float}
\usepackage{url}
\usepackage{booktabs}
\graphicspath{{./images/}}

\usepackage{tikz}
\usetikzlibrary{positioning, arrows.meta, fit, backgrounds, calc}

\usepackage{listings}
\usepackage{xcolor}
\usepackage{enumitem}
\usepackage{multirow}

\usepackage[table]{xcolor}
\definecolor{RedCell}{HTML}{D32F2F}
\definecolor{OrangeCell}{HTML}{F9A825}

\usepackage{caption}
\usepackage{rotating}

\definecolor{codegray}{rgb}{0.5,0.5,0.5}
\definecolor{codepurple}{rgb}{0.58,0,0.82}
\definecolor{backcolour}{rgb}{0.95,0.95,0.92}

\usepackage[most]{tcolorbox}
\usepackage{etoolbox} 

\definecolor{takeawaybg}{RGB}{244, 244, 255} 

\newcounter{takeawaycounter}
\tcbset{
  takeaway/.style={
    enhanced,
    colback=takeawaybg,
    colframe=black,
    boxrule=1.5pt,
    arc=4pt,
    left=2pt,
    right=2pt,
    top=2pt,
    bottom=2pt,
    fonttitle=\bfseries,
    before skip=4pt,
    after skip=4pt
  }
}

\newcounter{findingcounter}
\tcbset{
  finding/.style={
    enhanced,
    colback=takeawaybg,
    colframe=black,
    boxrule=1.5pt,
    arc=4pt,
    left=2pt,
    right=2pt,
    top=6pt,
    bottom=2pt,
    fonttitle=\bfseries,
    overlay={
      \node[fill=black, text=white, font=\bfseries, rounded corners=4pt, anchor=west, minimum height=15pt, minimum width=65pt] 
        at ([xshift=10pt,yshift=0pt]frame.north west) {Finding~\thefindingcounter};
    },
    before skip=12pt, after skip=5pt
  }
}

\newcommand{\smartbox}[2]{%
  \ifstrequal{#1}{takeaway}{%
    \refstepcounter{takeawaycounter}%
    \begin{tcolorbox}[takeaway]
      #2
    \end{tcolorbox}
  }{%
    \ifstrequal{#1}{finding}{%
      \refstepcounter{findingcounter}%
      \begin{tcolorbox}[finding]
        #2
      \end{tcolorbox}
    }{%
      \textcolor{red}{Unknown box type!}
    }%
  }%
}

\usepackage{hyperref}
\usepackage{xurl}
\usepackage{soul}
\hypersetup{
    breaklinks=true
}

\usepackage{pifont}

\ifCLASSOPTIONcompsoc
  \usepackage[nocompress]{cite}
\else
  \usepackage{cite}
\fi
\ifCLASSINFOpdf
\else
\fi

\begin{document}
%
\title{\textsc{A2ABreak}: Systematic Security Analysis of the A2A Protocol}

\maketitle

\begin{tikzpicture}[remember picture, overlay]
  \node[anchor=north, yshift=-0.5cm] at (current page.north) {
    \normalfont\small\textit{Paper accepted at 42nd IEEE Annual Computer Security Applications Conference (ACSAC '26)}
  };
\end{tikzpicture}

\begin{abstract}
The Agent2Agent (A2A) protocol, now governed by the Linux Foundation, is an open standard that enables autonomous AI agents to discover, authenticate with, and delegate tasks to one another across organizational boundaries.
Designed to complement the Model Context Protocol (MCP) for tool integration, A2A is rapidly emerging as the horizontal communication layer of the multi-agent ecosystem.
Yet the protocol's security has received no systematic analysis.
This paper presents \textsc{A2ABreak}, the first rigorous systematic security analysis of the A2A protocol.
We introduce a novel framework that utilizes an LLM-assisted extraction of a verified finite-state machine directly from the natural-language specification, producing a unified model of 37~states and 76~transitions from 929 formalized statements, and then systematically reasons over this model to discover protocol-level vulnerabilities through adversarial verification, under a full-compliance assumption.
Our analysis uncovers 11 new vulnerabilities, each exploitable by a specification-compliant adversary without requiring any implementation flaw.
Among the findings are cross-client context injection through unprotected context identifiers, credential harvesting via multi-hop identity loss in delegation chains, and data exfiltration through rogue agents advertising unattested capability claims.
\textsc{A2ABreak} achieves 73.3\% precision and 84.6\% F1 against manual expert review, while a zero-shot LLM baseline operating over the same specification produces zero confirmed findings, demonstrating that explicit formal grounding is essential for sound protocol security analysis.
\end{abstract}

%
\IEEEpeerreviewmaketitle





%

\input{sections/Introduction}
\input{sections/Background}

\input{sections/Overview}
\input{sections/Design}
\input{sections/Evaluation}
\input{sections/SecurityAnalaysis}
\input{sections/RelatedWork}
\input{sections/Conclusion}
\section*{Acknowledgments}

The work reported in this paper has been supported by NSF under grants 2229876, 2112471 and 2625398, National Artificial Intelligence Research Resource Pilot award 250336, the University of Texas System Rising STARs Award (No. 40071109), and the startup funding from the University of Texas at Dallas.

\bibliographystyle{IEEEtran}
\bibliography{references}
\newpage

\appendices
\input{sections/Appendix}

\end{document}

%% file: sections/Introduction.tex
\section{Introduction}\label{sec:intro}

The emergence of autonomous AI agents, systems capable of planning, reasoning, and executing multi-step tasks with minimal human oversight, has created a pressing need for standardized inter-agent communication.
While individual agents have grown increasingly capable through frameworks such as LangChain~\cite{langchain}, AutoGen~\cite{autogen}, and CrewAI~\cite{crewai}, their ability to collaborate across organizational and vendor boundaries has remained fundamentally ad hoc: each integration requires bespoke connectors, proprietary message formats, and manually negotiated trust assumptions.
The Agent2Agent (A2A) protocol, introduced by Google in April 2025, directly targets this gap by defining an open, HTTP-based standard for agent discovery, authentication, task delegation, and result exchange~\cite{a2a-spec}.
Within months of its release, A2A attracted endorsement from over 50 technology partners, including Salesforce, SAP, ServiceNow, MongoDB, and LangChain, and in June 2025, governance was transferred to the Linux Foundation as an open-source project~\cite{ibm_a2a, linuxa2a, organization}.\looseness=-1

A2A is explicitly designed to complement, rather than replace, the Model Context Protocol (MCP), which Anthropic released in November 2024 to standardize the interface between an agent and its local tools~\cite{anthropic2024mcp}.
Where MCP governs the \emph{vertical} relationship between an agent and its data sources, APIs, and function calls, A2A governs the \emph{horizontal} relationship between autonomous peers.
In a typical production deployment, an orchestrator agent uses MCP to query internal databases and invoke local tools, then uses A2A to delegate specialized subtasks, billing, compliance review, and document generation to remote agents operated by entirely separate organizations.
Together, the two protocols form the foundational communication stack for the emerging multi-agent ecosystem, and their joint adoption trajectory suggests that A2A-mediated interactions will soon be the basis of critical enterprise workflows in finance, healthcare, supply-chain management, and government operations.\looseness=-1

This trajectory, however, demands scrutiny.
A2A introduces a fundamentally different and substantially broader attack surface than prior agent integration paradigms.
The protocol's central design principle is \emph{opaque execution}: a client agent delegates a task to a remote agent without any visibility into the remote agent's internal reasoning, tool invocations, or intermediate state~\cite{a2a-spec}.
This opacity enables cross-vendor collaboration and protects intellectual property, but it also means that a client has no mechanism to verify what a remote agent actually does with a delegated task, the credentials it accumulates, or the artifacts it returns.
The security model compounds this exposure by relying predominantly on non-normative guidance for protective measures: the specification expresses critical security controls at the \texttt{SHOULD} or \texttt{MAY} level rather than as mandatory requirements, and leaves key mechanisms---including the sole primitive for verifying agent identity at discovery time---entirely optional.
The combination of opaque execution with this permissive security model creates a potentially large attack surface, whose boundaries have not been formally characterized.\looseness=-1

Despite the protocol's rapid adoption, the security properties of A2A have received virtually no systematic analysis in the research literature.
Existing work has either surveyed the protocol at a descriptive level alongside other agent interoperability standards~\cite{survey2025, anbiaee2026comparative} or proposed high-level security architectures for A2A-based systems without grounding them in a formal analysis of the protocol's state machine~\cite{habler2025building}.
Louck et al.~\cite{louck2025improving} addressed token-lifecycle privacy gaps without examining broader structural vulnerabilities.
More broadly, recent work has identified multi-agent interaction security as a fundamental open challenge, taxonomizing threats that emerge from inter-agent communication such as cascading prompt injection, credential harvesting, and coordinated manipulation~\cite{schroederdewitt2025open}, but no study has performed specification-level analysis of any individual protocol.
No prior work has systematically modeled the A2A interaction lifecycle, enumerated the protocol's attack surface from its specification, or constructed a formal model suitable for structured vulnerability discovery. Implementation-level techniques such as static analysis, fuzzing, or symbolic execution can test conformance, but cannot reveal specification gaps inherited by all compliant implementations. Specification-only analysis using large language models is also insufficient on its own: in our experiments, a zero-shot baseline with Claude \textsc{Opus}~4.6 and extended thinking produced nine candidate attacks across two independent runs, yet every candidate was rejected upon adversarial verification. Each attack trace violated normative requirements that the model failed to account for without an explicitly constructed formal model as a grounding mechanism.

\smallskip
\noindent\textbf{Challenges.}
Performing a rigorous, specification-driven security analysis of A2A requires overcoming three technical challenges.
First, natural-language protocol specifications do not define behavioral rules in a logically precise form; constraints, triggers, guard conditions, and state transitions are expressed in prose that lacks formal semantics, and the same sentence may simultaneously describe a precondition, a normative requirement, and a data-type constraint without syntactic distinction, making it infeasible to extract a faithful formal model through direct interpretation (\textbf{C1}).
Second, LLM-based generation in protocol analysis is inherently open-ended: outputs have no ground-truth reference model to validate against, and there is no mechanism to detect when a generated artifact lacks a normative basis or introduces inconsistencies.
Manual construction avoids this but is prohibitively expensive; prior work on cellular protocol specifications required approximately 2,800 person-hours of expert annotation~\cite{hermes} and cannot track rapid specification updates (\textbf{C2}).
Third, vulnerability analysis must be \emph{sound}: every reported finding must be grounded in genuine normative gaps, extracted rules must correspond to actual normative statements, formal representations must preserve their semantics, and the reported missing primitive must be genuinely absent from the specification rather than defined in a different section (\textbf{C3}).

\smallskip
\noindent\textbf{Our Approach.}
We present \textsc{A2ABreak}, a framework that combines LLM-assisted formal modeling with adversarial verification to systematically discover protocol-level vulnerabilities from natural-language specifications.
\textsc{A2ABreak} is organized into three stages.
Stage~A formalizes the natural-language specification into a verified corpus of 929 structured statements through dual-pass semantic separation: an independent structural extraction pass and a behavioral extraction pass, reconciled and verified against the specification before downstream use.
Stage~B constructs a unified finite-state machine (FSM) from this corpus, building six per-stage FSMs independently, resolving inter-stage transitions, and applying deterministic and semantic deduplication to produce a final model of 37~states and 76~transitions, a 92\% end-to-end reduction from the 929 extracted statements.
Stage~C reasons over the unified FSM to discover vulnerability candidates and subjects each to adversarial verification that grounds every attack trace in valid FSM states. It further confirms executability under full protocol compliance and verifies that the identified missing primitive is genuinely absent from the specification.
All LLM-assisted steps operate under a domain-specific constraint language that forces machine-verifiable formal representations and prevents free-text hallucination, while human checkpoints at stage boundaries ensure that errors do not propagate silently into downstream analysis.

\smallskip
\noindent\textbf{Findings.}
\textsc{A2ABreak} identifies eleven protocol-level vulnerabilities spanning discovery, initiation, task execution, and interruption, each exploitable under full specification compliance without requiring any implementation flaw or misconfiguration.
The analysis reveals that the specification's non-normative security posture leaves concrete, exploitable gaps across the protocol lifecycle.
Conversation contexts carry no ownership binding or access control (spec sections \S3.4.1, \S3.4.3), enabling an authenticated adversary to inject tasks into another client's context, access the victim's accumulated state, and poison subsequent interactions.
Identity is established exclusively at the transport layer and is not propagated across delegation hops (spec sections \S7, \S7.6.2), allowing a malicious intermediary to silently harvest forwarded credentials because downstream agents and credential providers have no means to verify the original principal.
The \texttt{AgentSkill} data model consists entirely of self-asserted fields with no verification or attestation mechanism (spec sections \S4.4.5, \S8.4), enabling a fully compliant rogue agent to advertise false capabilities, receive delegated tasks containing confidential data, and return fabricated artifacts without triggering any protocol-level error.
The analysis achieves a precision of 73.3\% and an F1 score of 84.6\% against manual expert review, substantially outperforming a zero-shot baseline that produces zero confirmed findings.

\smallskip
\noindent\textbf{Contributions.} This paper makes the following contributions:

\begin{itemize}[leftmargin=*, nosep]
    \item \textbf{LLM-assisted protocol analysis framework.} We design and implement \textsc{A2ABreak}, a three-stage framework that transforms natural-language protocol specifications into verified finite-state machines and systematically discovers protocol-level vulnerabilities through constrained LLM reasoning and adversarial verification.
    We validate FSM construction on all 14 protocols of PSMBench~\cite{NEURIPS2025_521bd958}. Under the benchmark's own evaluator, \textsc{A2ABreak} achieves the highest transition recall of all evaluated systems and ranks second in state F1; with semantic label matching, it recovers 104 of 108 states and 210 of 297 transitions. The framework also substantially outperforms zero-shot LLM analysis (Section~\ref{sec:eval}).\looseness=-1

    \item \textbf{Formal lifecycle model.} We construct the first finite-state machine model of the complete A2A protocol interaction lifecycle, decomposing it into six stage-specific FSMs that are merged into a unified model of 37~states and 76~transitions, capturing all states, transitions, and failure modes defined by the specification (Section~\ref{sec:design}).

    \item \textbf{Systematic vulnerability analysis.} We identify and validate eleven protocol-level vulnerabilities in A2A, mapped to specific FSM stages, specification sections, and STRIDE categories, providing the first structured threat model for the protocol.
    The analysis achieves a precision of 73.3\% and an F1 score of 84.6\% against expert manual review (Section~\ref{sec:sec_analysis}).

    \item \textbf{Open-source framework and artifacts.} We release \textsc{A2ABreak}'s full source code, all LLM prompts, stage-specific FSM models, and experimental artifacts to support reproducibility. \url{https://github.com/arlotfi79/A2ABreak}
\end{itemize}

\smallskip
\noindent\textbf{Responsible disclosure.} We reported our two most severe findings, Cross-Client Context Injection and Multi-Hop Identity Loss, together with proposed mitigations, to the A2A project maintainers under the Linux Foundation. The A2A Technical Steering Committee has responded, and we are working with them to address the issues in the specification. All artifacts of \textsc{A2ABreak} are available in our repository, including the extracted statement corpus, the unified FSM, the vulnerability analysis traces, the formal models, and the PSMBench outputs.

\smallskip
\noindent\textbf{Roadmap.}
Section~\ref{sec:background} provides background on MCP, A2A, and their complementary roles.
Section~\ref{sec:overview} defines the scope, threat model, and problem statement, and discusses the key technical challenges addressed by \textsc{A2ABreak}.
Section~\ref{sec:design} details the three-stage framework: specification formalization, FSM construction, and security analysis.
Section~\ref{sec:eval} evaluates the framework's correctness, refinement contribution, and cost.
Section~\ref{sec:sec_analysis} presents the eleven validated vulnerabilities with representative attack scenarios.
Section~\ref{sec:related} surveys related work, Section~\ref{sec:conclusion_and_future_work} concludes the paper and discusses future directions.

%% file: sections/Background.tex
\section{Background}
\label{sec:background}
In this section, we discuss the relevant details of A2A and MCP protocol and further highlight their distinction.

\subsection{Agent2Agent Protocol (A2A)}
\label{sec:bg:a2a}

The Agent2Agent (A2A) protocol is an open standard introduced by Google in
April 2025 to enable interoperability between autonomous AI agents across
frameworks, vendors, and organizational
boundaries~\cite{a2a-spec,ibm_a2a}. A2A defines a client--server model
in which a \emph{Client Agent} discovers, authenticates with, and delegates
tasks to a \emph{Remote Agent} over HTTPS. A central design principle is
\emph{opaque execution}: agents collaborate based on declared capabilities
and exchanged data, without exposing internal reasoning, memory, or tool
implementations to one another~\cite{a2a-spec}. The v1.0.0 specification
defines the data model normatively in Protocol Buffers and provides three
protocol bindings: JSON-RPC~2.0, gRPC, and HTTP+JSON/REST. The specification
uses RFC~2119 keywords~\cite{rfc2119} to express its requirements at varying
normative strengths. A2A launched with support from over 50
technology partners; governance was transferred to the Linux Foundation in
June 2025~\cite{platformeng2025,ibm_a2a}.

A complete A2A interaction proceeds through six stages that structure our
FSM model (Section~\ref{sec:design}) and vulnerability analysis
(Section~\ref{sec:sec_analysis}).
In \emph{Stage~1 (Discovery)}, the client fetches the \emph{AgentCard} a
self-asserted JSON document that declares the agent's identity, skills,
endpoint, and authentication requirements which serve as the sole trust
anchor for all subsequent interaction; signing is
optional~\cite{a2a-spec,survey2025}.
In \emph{Stage~2 (Authentication)}, the client authenticates using the
scheme declared in the AgentCard (OAuth~2.0, API keys, mTLS); all
credentials are carried at the transport layer, and authentication is
strictly hop-by-hop~\cite{a2a-spec,galileo2025}.
In \emph{Stage~3 (Initiation)}, the client sends an initial message and the agent responds with either a stateless reply or a stateful \emph{Task}
that enters the lifecycle.
In \emph{Stage~4 (Task Execution)}, the task is acknowledged and actively
processed, with updates delivered via polling, streaming, or push
notifications.
In \emph{Stage~5 (Interruption)}, the agent may pause execution to request
additional input or re-authentication; the client responds, and the task
returns to Stage~4. This optional loop may repeat unboundedly.
In \emph{Stage~6 (Termination)}, the task reaches a terminal
state: completed, failed, canceled, or rejected~\cite{a2a-spec}.

\subsection{Model Context Protocol (MCP)}
\label{sec:bg:mcp}
Introduced by Anthropic in November 2024~\cite{anthropic2024mcp}, the Model Context Protocol (MCP) standardizes the interface between an LLM application and its local tools and data sources via a client-server architecture over JSON-RPC~2.0~\cite{jsonrpc2,mcpspec2025,infomcp2024}. The protocol addresses the so-called \emph{$N\times
M$ integration problem}: without a shared standard, each combination of AI model and external tool requires a bespoke connector, causing integration complexity that grows quadratically with the number of models and services~\cite{anthropic2024mcp}. MCP reduces this to an \emph{N+M} architecture by defining a universal capability surface through three protocol primitives, namely, \textit{tools} (executable functions an agent can invoke), \textit{resources} (read-only data sources that supply context), and \textit{prompts} (reusable templates that structure model--server interaction); all discoverable at runtime via a machine-readable handshake~\cite{mcpspec2025}. MCP was adopted by OpenAI and Google DeepMind in early 2025~\cite{openaimcp,googlemcp}, and the November 2025 specification revision added support for asynchronous operations, stateless transports, and server identity, broadening its applicability to cloud-hosted and enterprise deployments. In December 2025, Anthropic donated the protocol to the Linux Foundation~\cite{wikipedia_mcp}, and by mid-2026, a release candidate for a further major revision, introducing a fully stateless protocol core and an extensions framework, was under community review.

\subsection{MCP vs.\ A2A}
\label{sec:bg:comparison}

\begin{figure}
  \centering
  \includegraphics[width=\columnwidth]{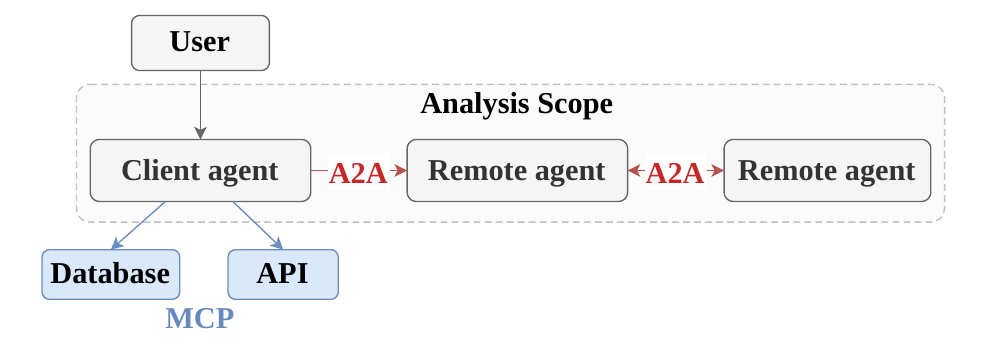}
  \captionsetup{justification=centering}
  \caption{MCP vs. A2A protocol scope}
  \vspace{-0.4cm}
  \label{fig:mcp_vs_a2a}
\end{figure}

MCP and A2A are complementary~\cite{ibm_a2a} (Figure~\ref{fig:mcp_vs_a2a}): MCP governs the
\emph{vertical} relationship between an agent and its tools, while A2A
governs the \emph{horizontal} relationship between autonomous peers. The
critical distinction lies in their trust models. 
MCP servers are subordinate
tool providers under full host control; A2A remote agents are opaque,
autonomous peers whose internal state is hidden by
design~\cite{a2a-spec}. This opacity prevents clients from verifying
what a remote agent does with delegated data, accumulated credentials, or
returned artifacts, introducing the broader attack surface analyzed in this paper.\looseness=-1
\vspace{-0.3cm}

%% file: sections/Overview.tex
\section{Overview}\label{sec:overview}
\vspace{-0.2cm}
In this section, we define the scope and threat model of our
analysis, formally state the problem, and discuss the critical challenges
tackled by \textsc{A2ABreak}.\looseness=-1
\vspace{-0.2cm}
\subsection{Scope}
\vspace{-0.2cm}
This work targets protocol-level security analysis of the A2A
specification v1.0.0~\cite{a2a-spec}. The analysis treats the protocol's
normative rules as the sole evidence base and determines whether
they are sufficient to guarantee key security properties, or whether
gaps permit attacks even under full compliance. We do not enumerate
all possible vulnerabilities, verify specific agent implementations
or address transport-layer security concerns.\looseness=-1
\vspace{-0.2cm}
\subsection{Threat Model}
\vspace{-0.1cm}
We consider an adversary who is computationally bounded and cannot
break standard cryptographic primitives but exploits gaps in the A2A
protocol specification by sending protocol-conformant messages. The
adversary has complete knowledge of the specification and operates
exclusively through A2A-defined operations. The adversary may act as a client initiating requests to a legitimate agent, as a server receiving
tasks from a legitimate client, or as an intermediary occupying both
roles within a multi-agent delegation chain. In all cases, the
adversary is fully authenticated and compliant; their advantage
derives entirely from what the protocol fails to define, not from
any deviation from the specification.

\noindent\textbf{Full compliance.} A party is \emph{fully compliant} if it satisfies all \textsc{must} and \textsc{should} requirements and enables all \textsc{may}-level safeguards, such as Agent Card signing. This excludes findings caused by omitted optional protections. Every reported vulnerability persists with all optional safeguards enabled, because no normative requirement prevents it.

\subsection{Problem Statement}

Given a protocol specification in natural language, \textsc{A2ABreak}
aims to generate a formal model and systematically discover
protocol-level vulnerabilities.

\noindent\textbf{Specification corpus.} Let
$\mathcal{D} = \{d_1, d_2, \ldots, d_M\}$ denote the set of the $M$
sections of the A2A specification. Each section $d_k$ is parsed into
a sequence of statements $\mathcal{C}_k = \langle c_1^{(k)},
c_2^{(k)}, \ldots \rangle$, where each statement is either a
structural definition or a normative rule expressed using RFC~2119
keywords. The full corpus is $\mathcal{C} = \bigcup_k \mathcal{C}_k$
with $|\mathcal{C}| = N$ after deduplication.\looseness=-1

\noindent\textbf{Protocol FSM.} From $\mathcal{C}$, we derive a
finite-state machine $\mathcal{F} = (S, s_0, T, \Sigma, S_f)$,
where $S$ is the set of protocol states, $s_0 \in S$ the initial
state, $S_f \subseteq S$ the terminal states, $\Sigma$ the event
alphabet, and $T \subseteq S \times \Sigma \times \mathcal{G} \times
S$ the transition relation with guard conditions $\mathcal{G}$. Each
transition $\tau = (s, \sigma, g, s') \in T$ represents a state
change from $s$ to $s'$ triggered by event $\sigma$ under guard $g$.\looseness=-1

\noindent\textbf{Protocol Vulnerability.} A protocol-level vulnerability is a
tuple $\mathcal{V} = (\pi, \rho, \phi)$: a feasible attack trace
$\pi = \langle \tau_1, \ldots, \tau_k \rangle$ through $\mathcal{F}$,
a violated security property $\rho$, and a missing primitive $\phi$
whose absence permits the violation. A trace is \emph{feasible} if
every transition exists in $T$ and every guard can be satisfied by a
fully compliant party. We classify $\mathcal{V}$ as a vulnerability
only if $\rho$ is a security goal stated by the specification and no
normative requirement, at any modality, prevents $\pi$.

\vspace{-0.2cm}
\subsection{Challenges}
\vspace{-0.1cm}
\noindent\textbf{C1: Infeasibility of semantic interpretation of
natural-language specifications.} Natural-language protocol
specifications do not define behavioral rules in a logically or
mathematically precise form. Constraints, triggers, guard conditions,
and state transitions are expressed in prose that lacks formal
semantics, i.e., the same sentence may simultaneously describe a
precondition, a normative requirement, and a data type constraint
without syntactic distinction. This ambiguity makes it infeasible to
extract a faithful formal model through direct interpretation.

\noindent\textbf{C2: Ensuring correctness of open-ended model
generation.} LLM-based generation in protocol analysis is inherently
open-ended: outputs have no formal reference model to validate
against, and there is no mechanism to detect when a generated artifact lacks a normative basis or introduces inconsistencies.
Standard models lack the sustained reasoning capacity to make 
interdependent decisions reliably over large inputs, and
errors propagate invisibly into downstream analysis. Manual
construction avoids this, but is prohibitively expensive. Prior work
on cellular protocols required approximately 2,800 man-hours of
expert annotation~\cite{hermes} and cannot track rapid
specification updates. Directly prompting an LLM to identify
vulnerabilities without a formal model produces plausible but incorrect findings. From our experiments, we see zero-shot baselines generate nine candidates across two runs, all rejected upon manual review.\looseness=-1

\noindent\textbf{C3: Ensuring soundness of automatic vulnerability
analysis.} 
For vulnerability analysis, soundness requires that every output of the framework is
faithful to the specification---extracted rules must correspond to
actual normative statements, formal representations must preserve
their semantics, and reported vulnerabilities must represent genuine
design flaws. Without explicit constraints on what the model can
produce at each stage, errors accumulate silently. For instance, an extraction stage may invent a guard condition that does not appear in the
specification, an FSM construction stage may connect states that the
protocol never links, and a vulnerability analysis stage may report a
primitive as absent when it is defined in a different section. Each
error is individually plausible, and downstream stages have no
mechanism to detect that their inputs are wrong.

\subsection{Our Solution Approaches}
\vspace{-0.2cm}
\noindent\textbf{S1: Dual-pass semantic separation.}
\textsc{A2ABreak} addresses the semantic interpretation challenge by
separating specification interpretation into two independent passes.
The key insight is that structural definitions and behavioral rules,
though interleaved in prose, impose fundamentally different extraction
constraints; a single pass conflates them, producing artifacts where
type definitions are misread as transitions and vice versa. The first
pass targets structural content: data types, field definitions, object
hierarchies, and enumerations. The second targets behavioral content:
state transitions, guard conditions, normative requirements, and
protocol operations. The two passes are executed independently, and
their outputs are reconciled before further processing, preventing
cross-contamination between definitional and behavioral semantics and
allowing each pass to apply an extraction schema optimized for its
specific content category.\looseness=-1

\noindent\textbf{S2: Input reduction and extended thinking.} 
\textsc{A2ABreak} addresses this challenge by decomposing
the analysis into a sequence of focused steps, each operating over a
narrow, task-specific input slice rather than the full specification.
The insight is that open-ended generation over large inputs is where
LLM reasoning is most error-prone; by restricting each step's scope,
we convert an unconstrained synthesis problem into a series of
bounded decisions that are individually verifiable. As a concrete
example, our framework begins with an extraction stage that parses the
specification into individual formal statements---929 statement in total---but only 283 (30\%) of these carry transition structure and
enter FSM construction; the remainder are filtered out before
downstream steps. Within each step, all generation is delegated to
a model with extended thinking capabilities, providing the sustained
reasoning depth needed to make interdependent decisions reliably over
the reduced inputs.

\noindent\textbf{S3: Domain-specific constraint language
for analysis soundness.} \textsc{A2ABreak} embeds formally specified
constraints across all framework prompts that collectively function as
a domain-specific language governing both extraction and analysis.
In the extraction stages, the constraints define a formal transition
schema ($s_1\_c/a\_\_s_2$) with typed fields for guards, triggers,
and post-conditions expressed in a logical notation
(\texttt{\&}, \texttt{|}, \texttt{!}, dot-qualified attributes),
forcing the model to produce machine-verifiable formal
representations rather than free-text summaries. In the analysis
stages, the constraints define what counts as a valid finding: the
flaw must exist in what the protocol does not say; vague delegation
is not intentional exclusion; and an outcome mandate without a
mechanism is a missing primitive. The verification stage enforces
these through four adversarial steps: ground the trace in FSM states,
confirm executability under full compliance, verify no normative rule
at any modality prevents the trace, and confirm the primitive is
genuinely absent. Candidates that fail any step are discarded with
specific falsifying evidence.

%% file: sections/Design.tex
\vspace{-0.2cm}
\section{Design of \textsc{A2ABreak}}\label{sec:design}
\begin{figure*}[t]
  \centering
  \includegraphics[width=0.25\textwidth,trim = 6.5cm 0cm 6.5cm 0cm, clip, angle=270]{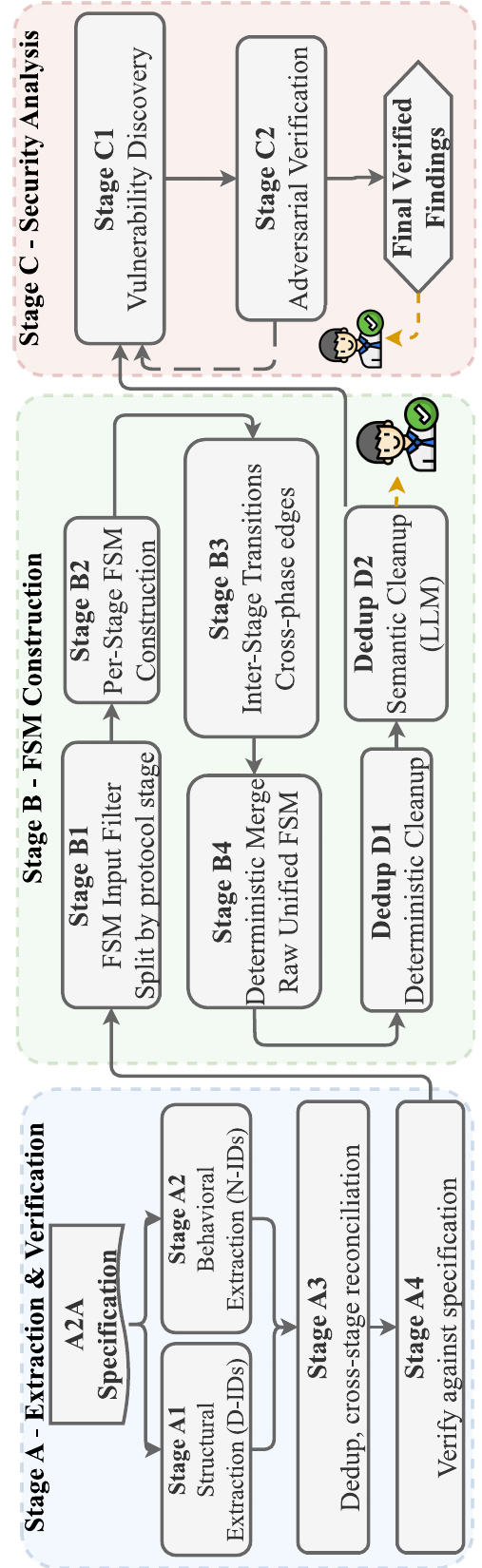}
  \captionsetup{justification=centering,font=footnotesize}
  \vspace{-0.2cm}
  \caption{Overview of the A2ABreak.}
  \label{fig:pipeline_diagram}
  \vspace{-.4cm}
\end{figure*}

\vspace{-0.2cm}
\textsc{A2ABreak} has three stages (Figure~\ref{fig:pipeline_diagram}). Stage A converts the natural-language specification into verified structured statements, Stage B builds a unified FSM, and Stage C analyzes it for protocol-level vulnerabilities. Human checkpoints between stages prevent errors from propagating into the security analysis.

\vspace{-0.2cm}

\subsection{Stage A: Specification Formalization}
\vspace{-0.3cm}
Stage A transforms the A2A protocol specification from natural language into
a machine-processable corpus of formal statements. Because the specification
mixes definitional structure with RFC 2119 behavioral rules, extraction is
split into two independent passes targeting these orthogonal concerns
separately. Every extracted statement is encoded as a component of the FSM
tuple $\mathcal{F} = (S, s_0, T, \Sigma, S_f)$: states populate $S$,
transitions populate $T$, and trigger strings contribute to the event
alphabet $\Sigma$. Each transition is represented as a tuple
$\langle s_1, \text{pre\_cond}, \text{trigger}, \text{post\_cond}, s_2
\rangle$ instantiating the formal notation $s_1 \xrightarrow{c/a} s_2$,
where $c$ is the guard condition and $a$ is the resulting effect.

\noindent\textit{\ul{A1: Structural Extraction.}}
Step A1 extracts the implicit formal structure that all subsequent behavioral
analysis depends on. It produces three categories of statement:
\textsc{FSM\_State} entries name every distinct lifecycle condition a protocol
entity can occupy and classify each as initial, intermediate, or terminal;
\textsc{Constraint} entries capture global data invariants that hold across
all states; and \textsc{Implicit\_Transition} entries encode state changes
implied by definition tables and protocol flow diagrams, even when no RFC
2119 keyword accompanies them. Each statement receives a stable identifier
and is annotated with the protocol stage it belongs to.

\noindent\textit{\ul{A2: Behavioral Extraction.}}
Step A2 performs an independent pass over the same source, targeting
exclusively sentences that contain an RFC 2119 keyword. Each such sentence is
encoded as a \textsc{Behavioral} transition statement using the identical
schema used by Step A1, preserving the normative obligation strength
(\texttt{must}, \texttt{shall}, \texttt{should}, \texttt{may}) for downstream
analysis. 
Step A2 deliberately does not re-extract definition tables or
enumeration values, keeping the two corpora semantically disjoint.

\noindent\textit{\ul{A3: Reconciliation.}}
The outputs of Steps A1 and A2 are merged into a single unified statement
list. Exact and near-duplicate statements arising from the same normative
rule being captured by both passes are detected and collapsed into a
canonical form, preserving source identifiers from both origins for
traceability. Ambiguous cases are resolved by an LLM pass and flagged for
mandatory human review before \textsc{A2ABreak} advances.

\noindent\textit{\ul{A4: Specification Verification.}}
Every statement in the reconciled corpus is independently verified against
the live specification source text before any FSM construction begins. Each
statement's extracted fields are confirmed against the originating
specification section, with automated checks detecting hallucinated states,
misattributed normative keywords, and guard conditions absent from the source.
Statements that fail verification are corrected or removed, and the outcome
is recorded in a verification report that forms a mandatory human checkpoint
before Stage B.

\subsection{Stage B: FSM Construction}
\vspace{-0.2cm}
Stage B converts the verified statement corpus into a single unified FSM.
To manage the complexity of a six-stage protocol, per-stage FSMs are
constructed independently and then connected by resolving cross-stage
transitions before a final deterministic merge.

\noindent\textit{\ul{B1: FSM Input Filtering.}}
Step B1 is a deterministic, LLM-free filter that partitions the verified
corpus by protocol stage and retains only statements that carry sufficient
transition structure to contribute a state or edge to the graph. Its output
is six per-stage input bundles, one for each protocol stage.

\noindent\textit{\ul{B2: Per-Stage FSM Synthesis.}}
Step B2 constructs a complete FSM for each of the six protocol stages from
its B1 input bundle. For each stage, the synthesizer deduplicates states,
discards entries that are not protocol session states, and infers implicit
states referenced in transitions but absent from explicit definitions. Every
state is assigned an \texttt{actor} field, \texttt{client}, \texttt{server},
\texttt{shared}, or \texttt{unknown}, derived by reasoning about which
principal controls entry into and exit from that state. Shared-actor states
mark trust boundaries and are primary targets for the security analysis in
Stage C. Where the input statements leave gaps, the synthesizer queries the
live specification before recording unresolvable cases in diagnostic lists
that form a mandatory human checkpoint.

\noindent\textit{\ul{B3: Inter-Stage Transition Resolution.}}
Once all six per-stage FSMs exist, Step B3 identifies the cross-stage edges
that connect them. It examines exit points in each stage of the FSM and determines
which entry points of the downstream stage they connect to, along with the
trigger event and guard condition governing each handoff. The seven-stage boundaries resolved include the bidirectional cycle between task\_execution
and interruption. Step B3 produces only inter-stage transitions and does not
modify the intra-stage graphs from B2.

\noindent\textit{\ul{B4: Unified FSM Assembly.}}
Step B4 deterministically assembles the unified FSM by combining all six
per-stage FSMs from Step B2 and the inter-stage edges from Step B3 into a
single connected graph. The resulting unified FSM is then refined through two
deduplication sub-passes. Sub-pass~D1 is a deterministic workflow
that normalizes naming inconsistencies, merges states duplicated across
stages, removes identity inter-stage transitions that are artifacts of
stage-scoped modeling, and merges semantically equivalent duplicates
transitions. It concludes with structural validation and uses a heuristic
detectors to flag candidate issues, meta-states, absence states, and
unreachable non-protocol states for the subsequent pass. Sub-pass~D2 is a
targeted LLM pass that triages these candidates: removing abstract meta-states
and redistributing their behavior, collapsing trivial pass-through states,
discarding deployment-lifecycle states that no protocol operation can reach,
and adding missing incoming transitions for disconnected sub-lifecycles by
consulting the specification provenance attached to each state. The output is
a fully connected FSM with zero orphaned states and full referential integrity.
Figure~\ref{fig:unified_fsm} (available in Appendix~\ref{appendix:fsm}) illustrates the derived unified FSM for the A2A protocol, comprising 37 states and 76 transitions. For better resolution, we refer to the artifact repository.\looseness=-1


\subsection{Stage C: Security Analysis}
\vspace{-0.2cm}
Stage C reasons over the unified FSM to identify and validate protocol-level
security vulnerabilities. The analysis operates under a full compliance
assumption (Section~\ref{sec:overview}): all parties honor every normative
statement and enable every optional safeguard, implementations are correct,
and transport security is properly established.

\noindent\textit{\ul{C1: Vulnerability Discovery.}}
Step C1 examines the unified FSM for structural properties associated with
known security vulnerabilities: missing authorization guards on sensitive
transitions, absent checks at stage boundaries, states reachable without
traversing required predecessors, and event-alphabet gaps that permit
unanticipated message sequences. Before reporting any candidate, two filters
are applied. First, the attack trace must be executable entirely within the
A2A protocol as defined, with no out-of-band mechanisms. Second, a web search of the original specification must confirm that the identified gap is a genuine
missing primitive rather than an intentional design choice, vague delegation
to implementations does not qualify as intentional exclusion. C1 is
executed multiple times as independent discovery passes; candidates from all
runs are merged before Step C2, improving recall across the candidate set.

\noindent\textit{\ul{C2: Adversarial and Manual Verification.}}
Step C2 subjects every C1 candidate to a combined automated and human
verification process. Each candidate is first grounded in the unified FSM,
discarding any attack path that cannot be reconstructed from valid states
and transitions. The remaining candidates are executed step by step under
full protocol compliance to confirm exploitability and to ensure that no
existing normative rule prevents the attack. The specification is then
searched comprehensively to verify that the identified primitive is truly
absent rather than defined elsewhere.

Confirmed candidates subsequently undergo manual review by the first three authors against the specification and unified FSM.
The reviewers validate the FSM trace and confirm that the cited specification evidence supports the claimed
absence of the missing primitive. Aided by the pipeline's citations, each finding took about 15 person-minutes or less to verify. 
Candidates that fail either automated
or manual scrutiny are downgraded to inconclusive or discarded with recorded
justification. The final output partitions the candidate set into confirmed
findings, discarded candidates, and inconclusive cases, providing a complete
audit trail from discovery through validation.

%% file: sections/Evaluation.tex
\section{Evaluation}
\label{sec:eval}

We evaluate the effectiveness of \textsc{A2ABreak} through the following research questions:
\begin{itemize} 
    \item \textbf{RQ1.} How does \textsc{A2ABreak} compare against a zero-shot analysis of the protocol?
    \item \textbf{RQ2.} How is the correctness of \textsc{A2ABreak} evaluated?
    \item \textbf{RQ3.} How does each pipeline stage contribute to FSM refinement?
    \item \textbf{RQ4.} Where does specification complexity concentrate across A2A protocol stages?\looseness=-1
    \item \textbf{RQ5.} What is the execution cost of \textsc{A2ABreak}?
\end{itemize} 

\noindent\textit{\ul{RQ1: Zero-Shot Comparison.}} 
To compare \textsc{A2ABreak} against a zero-shot baseline, we prompted
Claude Opus 4.6 with high reasoning enabled using only the A2A specification
and instructed it to identify protocol-level attack scenarios under the same
constraints enforced by our pipeline. Two independent zero-shot runs produced
nine candidate attacks. Applying the same Stage~C2
adversarial verification, all nine candidates were rejected, primarily because
their attack traces were already prohibited by normative requirements that the
model failed to account for. These results suggest that, without an explicitly
constructed FSM as a grounding mechanism, even advanced reasoning models lack
a precise representation of protocol behavior, limiting their ability to
distinguish genuine protocol design gaps from behaviors already constrained by
the specification.

\noindent\textit{\ul{RQ2: Correctness Validation.}}
We evaluate correctness at two levels: FSM construction (Stages~A--B) and vulnerability analysis (Stages~C1--C2). For FSM construction, we run the pipeline on all 14 PSMBench protocols~\cite{NEURIPS2025_521bd958}, and score its output with the benchmark's evaluator. Because the evaluator matches labels as strings, equivalent names such as \texttt{ESTAB} and \texttt{ESTABLISHED} count as mismatches. We therefore add an LLM judge that matches each ground-truth element to an extracted state or transition with the same protocol semantics. The judge neither modifies the FSM nor feeds decisions back into extraction. We only rename matched labels to the ground-truth wording and rescore, while unmatched edges still reduce precision.
Across the 14 protocols (macro-averaged), \textsc{A2ABreak} achieves 0.746 state F1 and transition precision, recall, and F1 of 0.318, 0.701, and 0.408. On TCP, it recovers all 11 states and 20 transitions. Precision remains limited because the pipeline exports all per-state behaviors defined by the specification, yielding 804 edges versus 297 in the diagram-level ground truth. Since judging is applied only to our output, baseline comparisons use the benchmark's original string matcher for all ten systems. Under this comparison, \textsc{A2ABreak} ranks second in state F1 at 0.714 and achieves the highest transition recall at 0.481, versus 0.415 for the strongest baseline. Table~\ref{tab:psmbench} in Appendix~\ref{appendix:psm_bench} reports both results (S = states, T = transitions).

For vulnerability analysis, across five analysis runs Stage~C1
produced 17 unique candidate vulnerabilities. Stage~C2 accepted 16 and correctly rejected~1. Manual review by the first three authors, 
applying the same compliance assumptions and evaluation
criteria used throughout the pipeline, validated 11 of the 16 as
true positives, identified 1~duplicate, and overturned 4~as false
positives (preventable by existing normative rules under
full compliance). Overall, \textsc{A2ABreak} achieved a precision of
73.3\% (11~TP out of 15~non-duplicate candidates), and an F1~score of 84.6\%.

\begin{figure}[h!]
  \centering
  \includegraphics[width=\columnwidth]{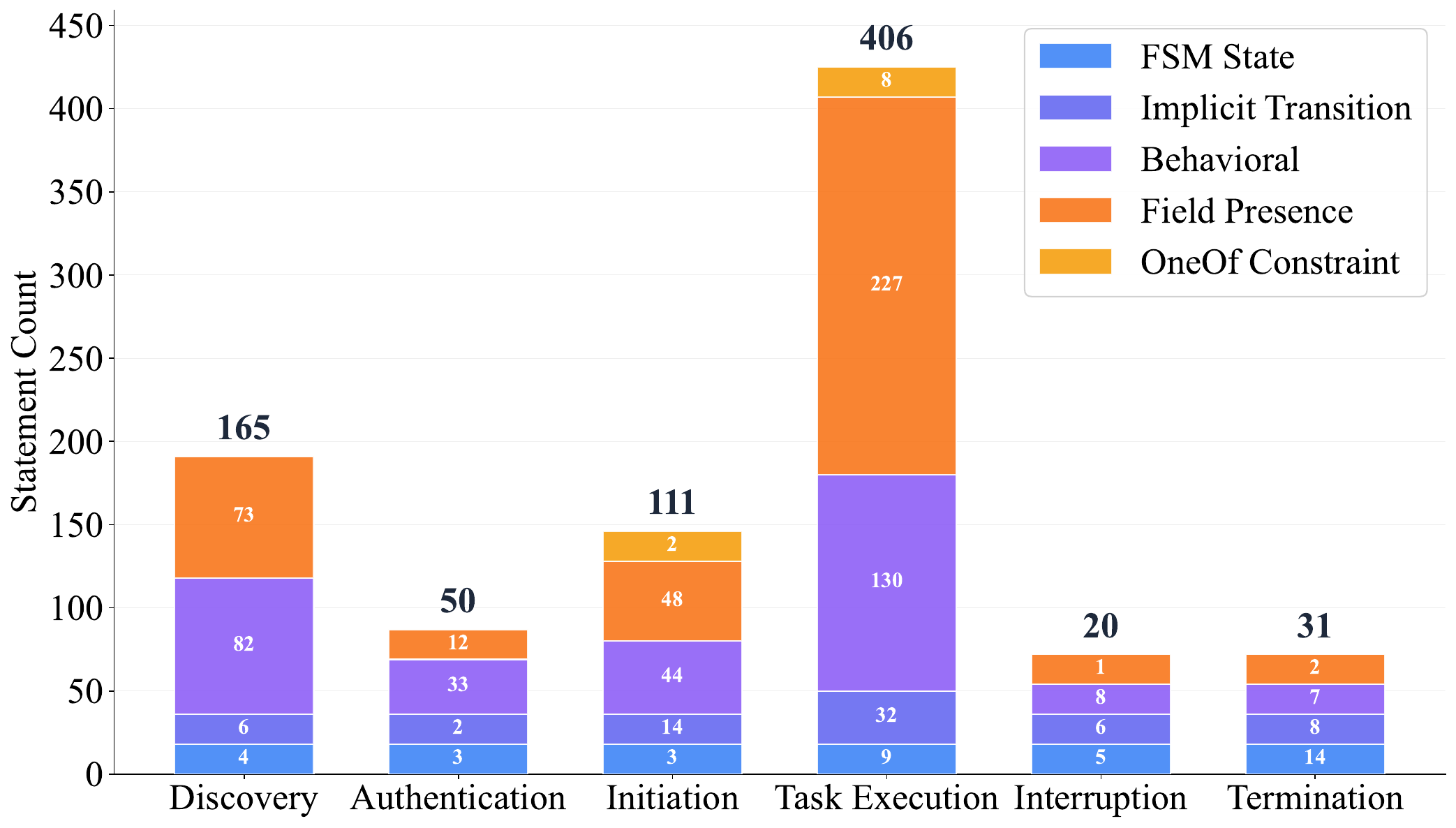}
  \captionsetup{justification=centering}
  \caption{Distribution of 929 verified statements.}
  \vspace{-.25cm}
  \label{fig:category_by_stage}
\end{figure}

\noindent\textit{\ul{RQ3: Pipeline Refinement.}} 
Figure~\ref{fig:category_by_stage} shows the 929 verified statements produced by Stage A, broken down by category and protocol stage. The two extraction passes yield five statement categories: FSM States name discrete lifecycle conditions an entity can occupy (e.g., \texttt{submitted}, \texttt{working}); Implicit Transitions capture state changes implied by prose without RFC 2119 keywords; Behavioral statements are explicit normative rules containing RFC 2119 keywords (MUST, SHOULD, MAY); Field Presence constraints record required or optional fields on protocol objects; and OneOf Constraints capture mutually exclusive field groups where exactly one must be present (e.g., a \texttt{Part} must contain one of \texttt{TextPart}, \texttt{FilePart}, or \texttt{DataPart}). Of these, only the first three carry transition structure and enter FSM construction: FSM States populate the state set $\mathcal{S}$, while Implicit Transitions and Behavioral statements populate the transition relation $\mathcal{T}$. Field Presence and OneOf Constraints, which dominate Discovery (73 of 165) and Task Execution (235 of 406), encode structural invariants but contribute no states or edges, and are filtered out before FSM synthesis.

\begin{figure}[h!]
  \centering
  \includegraphics[width=\columnwidth]{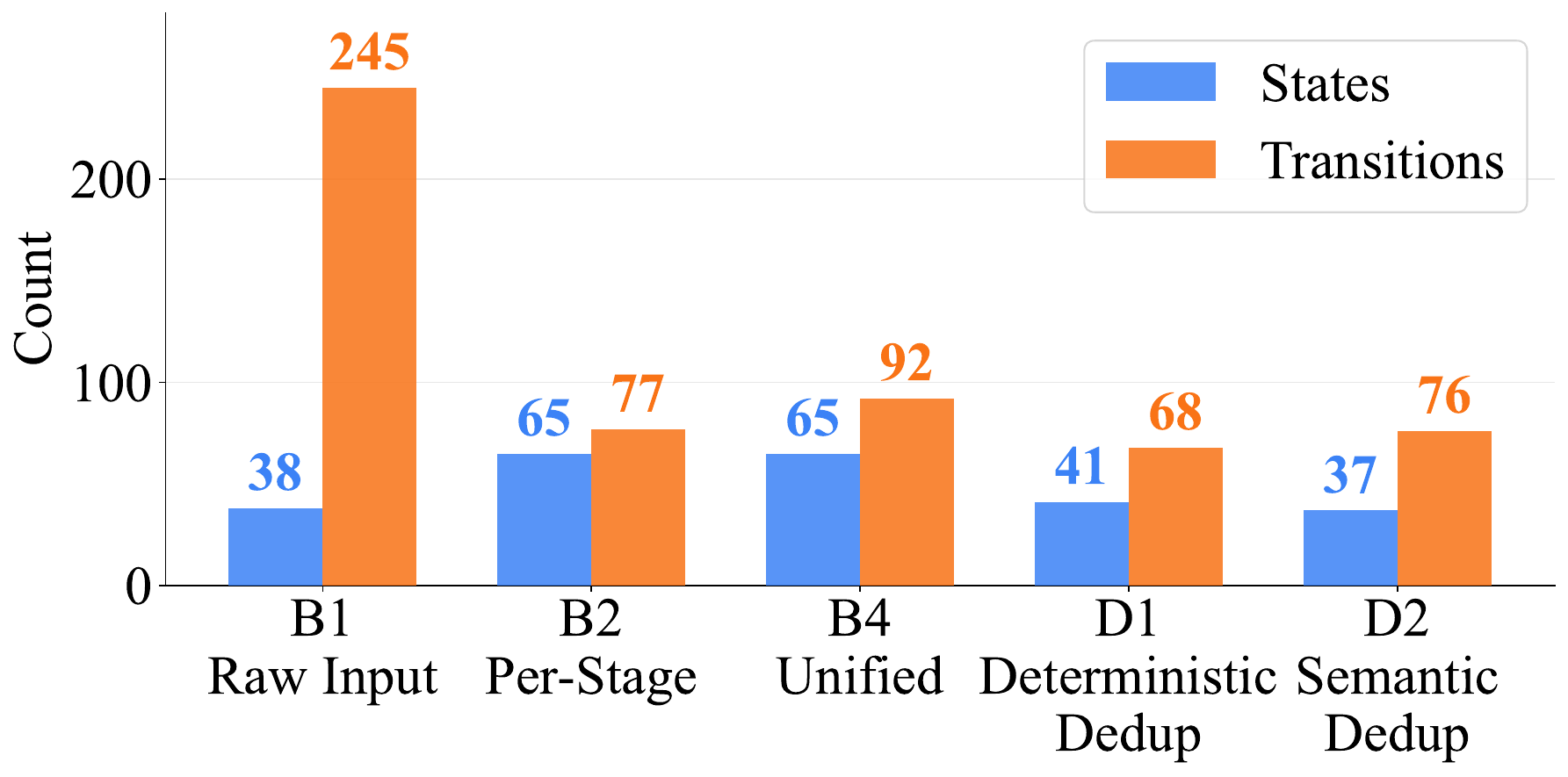}
  \captionsetup{justification=centering}
  \caption{FSM counts across \textsc{A2ABreak} stages.}
  \label{fig:fsm_waterfall}
\end{figure}

Figure~\ref{fig:fsm_waterfall} traces the subsequent FSM refinement. Of the 929 verified statements, only 283 (30\%) carry transition structure and enter FSM construction, producing a raw graph of 38 states and 245 transitions. Each subsequent stage targets a distinct class of redundancy: per-stage synthesis (B2) eliminates 168 intra-stage duplicate transitions, deterministic cleanup (D1) collapses 24 mechanically duplicate states across stage boundaries, and semantic deduplication (D2) resolves 4 naming-level equivalences detectable only by an LLM pass while recovering 8 previously obscured transitions. The final unified FSM of 37 states and 76 transitions represents a 69\% transition reduction from raw synthesis and a 92\% end-to-end reduction from the 929 extracted statements---confirming that each pipeline stage addresses an orthogonal source of noise whose removal is necessary for a tractable and faithful protocol model.

\begin{figure}[h!]
  \centering
  \includegraphics[width=\columnwidth]{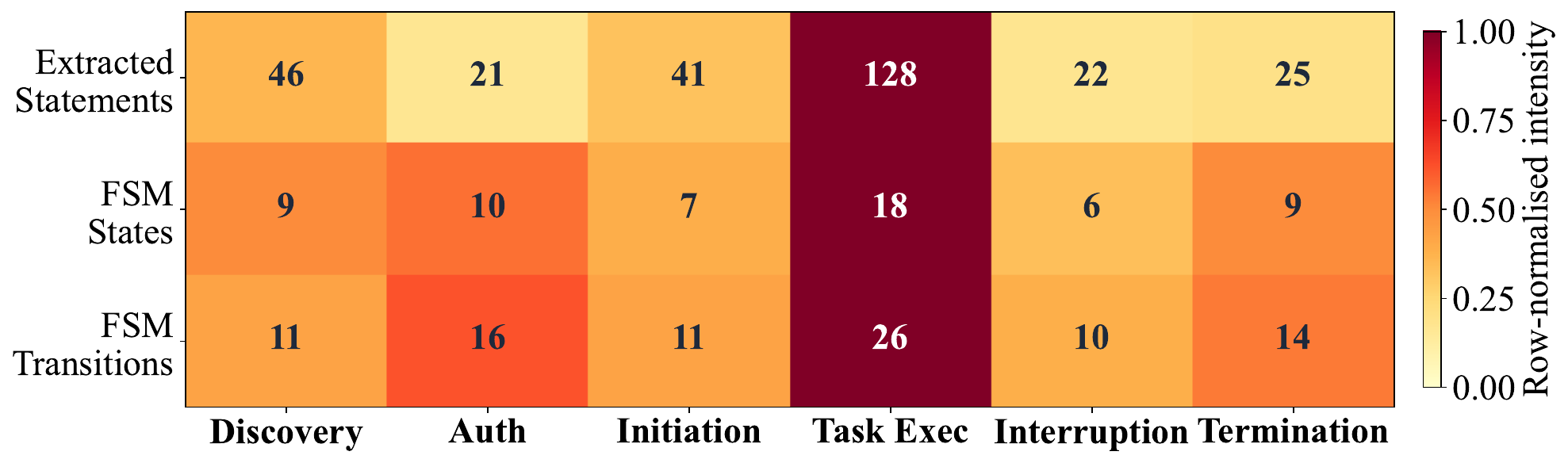}
  \captionsetup{justification=centering}
  \caption{Specification complexity of A2A protocol stages.}
  \label{fig:stage_heatmap}
\end{figure}

\noindent\textit{\ul{RQ4: Specification Complexity Concentration.}} 
Figure~\ref{fig:stage_heatmap} shows the distribution of extracted statements, FSM states, and FSM transitions across the six protocol stages after full deduplication. The pipeline reveals that textual volume is a misleading proxy for behavioral complexity: Authentication contains only 21 statements but produces 16 transitions (0.76 per statement), three times the ratio of Discovery (0.24). Interruption is the smallest stage by every metric yet encodes the re-entry loop central to credential accumulation in multi-hop chains. These mismatches are invisible from a linear reading of the specification and unrecoverable by monolithic FSM construction; they emerge precisely because the pipeline decomposes extraction by protocol stage, enabling quantitative cross-stage comparison that directs security analysis toward stages where behavioral density, not page count, indicates attack surface concentration.

\noindent\textit{\ul{RQ5: Execution Cost.}} 
Table~\ref{tab:cost} reports the API cost of a single end-to-end run of
\textsc{A2ABreak}. Stages A1--A4 use Claude Sonnet 4.6 with high reasoning,
while the more structurally complex Stages B2, B3, D2, C1, and C2 use Claude
Opus 4.6 with high reasoning. The total one-pass cost is \$40.97, of which
\$34.67 is consumed by Stages A and B, with Stage B2 being the single most
expensive step at \$11.64, followed by Stage A1 at \$9.04 and Stage A4 at
\$5.42. The FSM deduplication pass (D2) adds \$1.08, and a single Stage C
discovery and verification pass adds \$6.30. Mechanical stages involving no
LLM calls (B1, B4, D1) incur zero cost.

\begin{table}[h!]
\centering
\caption{Per-stage API cost for one \textsc{A2ABreak} run}
\label{tab:cost}
\begin{tabular}{p{0.08\columnwidth} p{0.22\columnwidth} p{0.18\columnwidth} p{0.18\columnwidth}}
\toprule
\textbf{Stage} & \textbf{Model} & \textbf{Reasoning} & \textbf{Cost (USD)} \\
\midrule
A1 & Sonnet 4.6 & High & \$9.04  \\
A2 & Sonnet 4.6 & High & \$5.46  \\
A3 & Sonnet 4.6 & High & \$0.38  \\
A4 & Sonnet 4.6 & High & \$5.42  \\
B2 & Opus 4.6   & High & \$11.64 \\
B3 & Opus 4.6   & High & \$1.65  \\
D2 & Opus 4.6   & High & \$1.08  \\
C1 & Opus 4.6   & High & \$3.20  \\
C2 & Opus 4.6   & High & \$3.10  \\
\midrule
\textbf{Total} & & & \textbf{\$40.97} \\
\bottomrule
\end{tabular}
\end{table}

%% file: sections/SecurityAnalaysis.tex
\vspace{-0.5cm}
\section{Security Analysis}\label{sec:sec_analysis}
To demonstrate that the vulnerabilities identified by \textsc{A2ABreak} carry concrete security
consequences, we present the full set of eleven validated findings in Table~\ref{tab:a2a-findings},
each exploitable under full specification compliance without requiring any implementation flaw
or misconfiguration. In this section, we discuss three representative findings in detail spanning
discovery, initiation, and interruption to illustrate the attack scenarios, message flows,
and impact of the missing protocol primitives.

\begin{table*}[t]
\centering
\scriptsize
\caption{\footnotesize Validated A2A Protocol-Level Vulnerabilities}
\vspace{-0.2cm}
\label{tab:a2a-findings}
\renewcommand{\arraystretch}{1.35}
\setlength{\tabcolsep}{5pt}
\begin{tabular}{l c p{0.22\textwidth} p{0.21\textwidth} p{0.19\textwidth} c}
\hline
\textbf{Stage} & \textbf{\#} & \textbf{Vulnerability} & \textbf{STRIDE Category} & \textbf{Impact (CIA)} & \textbf{Spec Reference} \\
\hline\hline
\multirow{2}{*}{\textit{\textbf{Discovery}}}
  & 1 & Unattested Skill Claims        & Spoofing                        & Confidentiality, Integrity      & \S4.4.5, \S8.4 \\
  & 2 & JWS Key Trust Model Gap        & Spoofing                        & Confidentiality                 & \S8.4 \\
\hline
\multirow{1}{*}{\textit{\textbf{Initiation}}}
  & 3 & Cross-Client Context Injection & Info.\ Disclosure, Tampering    & Confidentiality, Integrity      & \S3.4.1, \S3.4.3 \\
\hline
\multirow{4}{*}{\textit{\textbf{\shortstack{Task\\Execution}}}}
  & 4 & SSE Post-Revocation Leakage    & Info.\ Disclosure               & Confidentiality                 & \S7.4, \S3.2.3 \\
  & 5 & Artifact Chunk Integrity Gap   & Tampering                       & Integrity                       & \S4.2.2, \S3.1.6 \\
  & 6 & Concurrent Access TOCTOU       & Tampering, Elevation of Privilege  & Integrity, Confidentiality      & \S3.1.1, \S4.1 \\
  & 7 & Unverified Webhook URL         & DoS, Info.\ Disclosure          & Availability, Confidentiality   & \S13.2 \\
\hline
\multirow{4}{*}{\textit{\textbf{Interruption}}}
  & 8  & Auth Scope Amplification          & Elevation of Privilege          & Confidentiality, Integrity      & \S7.6.1, \S7.6.2 \\
  & 9  & Multi-Hop Identity Loss            & Spoofing, Info.\ Disclosure     & Confidentiality                 & \S7, \S7.6.2 \\
  & 10 & Circular Delegation Deadlock       & DoS                             & Availability                    & \S7.6.2, \S4.1.3 \\
  & 11 & No Timeout from Interrupted States & DoS                             & Availability                    & \S4.1.3 \\
\hline
\end{tabular}
\end{table*}

\noindent\textbf{(1) Cross-Client Context Injection.}
The A2A protocol allows agents to maintain persistent conversation
state across tasks via a shared \texttt{contextId}, accumulating
sensitive information such as user preferences, prior approvals, and
business data. However, the protocol defines no ownership model for
contexts: unlike tasks, which are bound to an authenticated principal
at creation, a \texttt{contextId} carries no creator field, no access
token, and no authorization requirement. Any authenticated client that
knows a valid \texttt{contextId} can inject a new task into that
context and receive a response informed by another client's
conversational state.\looseness=-1

\noindent\textit{\ul{Root cause.}}
The root cause is the specification permitting clients to supply a
\texttt{contextId} without ownership verification. Specification Section~\S3.4.1
allows servers to accept client-provided \texttt{contextId} values,
and \S3.4.3 defines cross-client context entry as a normative pattern:
clients \textsc{may} supply a \texttt{contextId} without a
\texttt{taskId} to start a new task within an existing context. The
\S13.1 authorization \textsc{must}s are scoped exclusively to task
operations, with no corresponding requirement at the context layer.

\begin{figure}[h!]
  \centering
  \includegraphics[width=0.6\columnwidth,  trim = 4cm 4cm 4cm 4cm, clip, angle=270]{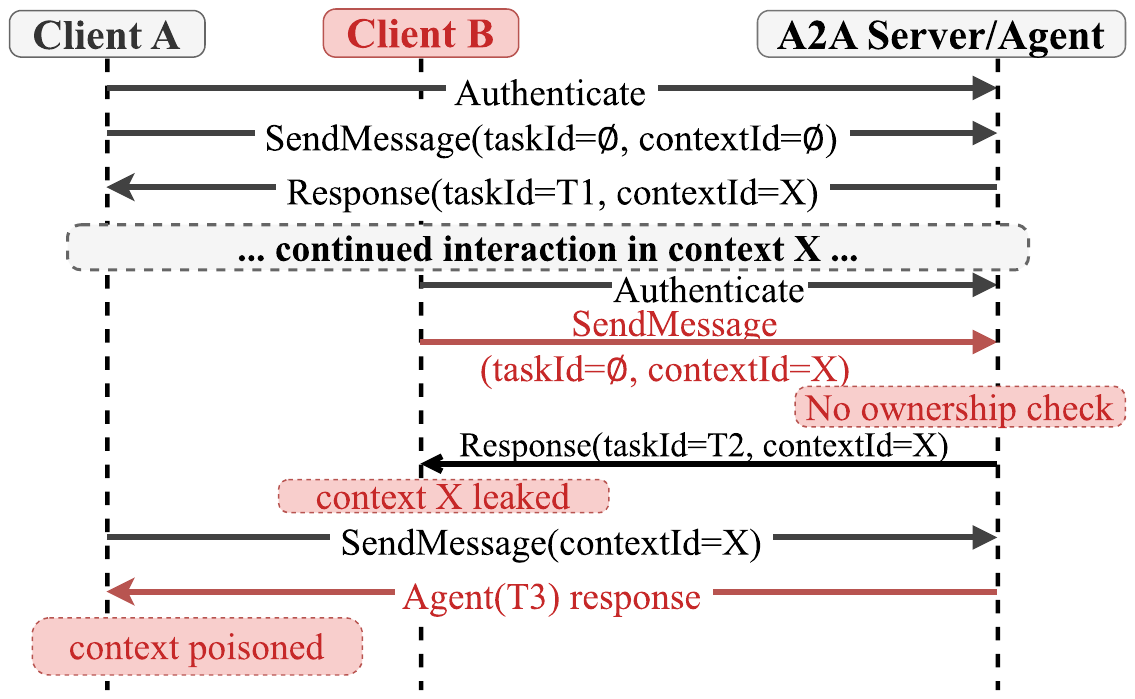}
  \captionsetup{justification=centering}
  \caption{Cross-Client Context Injection}
  \label{fig:vul:cross_client}
  \vspace{-.5cm}
\end{figure}

\noindent\textit{\ul{Attack.}}
The adversary (Client~B) first authenticates as a legitimate client.
As shown in Figure~\ref{fig:vul:cross_client}, Client~B then issues
\texttt{SendMessage(taskId=\(\emptyset\), contextId=X)}, where
\texttt{contextId=X} belongs to an ongoing session established by the
victim (Client~A). The server validates only
\texttt{contextId}--\texttt{taskId} consistency, a check that
passes and creates a new task in context \texttt{X} with no
ownership verification. The server then processes Client~B's task
against Client~A's accumulated conversational history and returns a
response informed by that state to Client~B.

\noindent\textit{\ul{Impact.}}
The attack produces two consequences. First, the adversary receives
responses informed by the victim's prior state, leaking sensitive
context, including approvals, and business data. Second, the
victim's subsequent tasks are processed against a history poisoned by
the adversary's injected messages, corrupting future results without
bypassing any task-level control. The adversary can repeatedly inject
into the same context, deepening the poisoning effect over time.

\noindent\textit{\ul{Mitigation.}}
The specification should define context ownership. Each \texttt{contextId} is bound server-side to its authenticated creator, mirroring task scoping in \S13.1. A message using a context owned by neither the caller nor an authorized participant \textsc{must} be rejected with the \S3.3.2 authorization error and \textsc{must not} access that context's state. \S3.4.1 should make this ownership check explicit, while legitimate multi-party contexts require owner-authorized sharing.

\noindent\textbf{(2) Multi-Hop Identity Loss.}
The A2A protocol enables multi-agent execution by delegating tasks across a
chain of agents. Identity, however, is established exclusively at the
transport layer and is not propagated across hops: each agent knows only its
immediate caller's identity. While sufficient for point-to-point interactions,
this design is architecturally incompatible with multi-hop delegation chains,
where downstream agents and credential providers must verify the original
principal's identity and delegation provenance.

\noindent\textit{\ul{Root cause.}}
The specification states that payloads do not carry user or client identity
directly and that identity is established at the transport layer~(spec section \S7).
No delegation context token, principal identity field, or chain-of-custody
primitive exists to bridge this hop-by-hop model with the multi-hop delegation
chains the protocol explicitly enables via
\texttt{TASK\_STATE\_AUTH\_REQUIRED}~(spec section~\S7.6.2).\looseness=-1

\begin{figure}[ht]
  \centering
  \includegraphics[width=0.55\columnwidth, trim = 3.5cm 1.3cm 3cm 1cm, clip, angle=270]{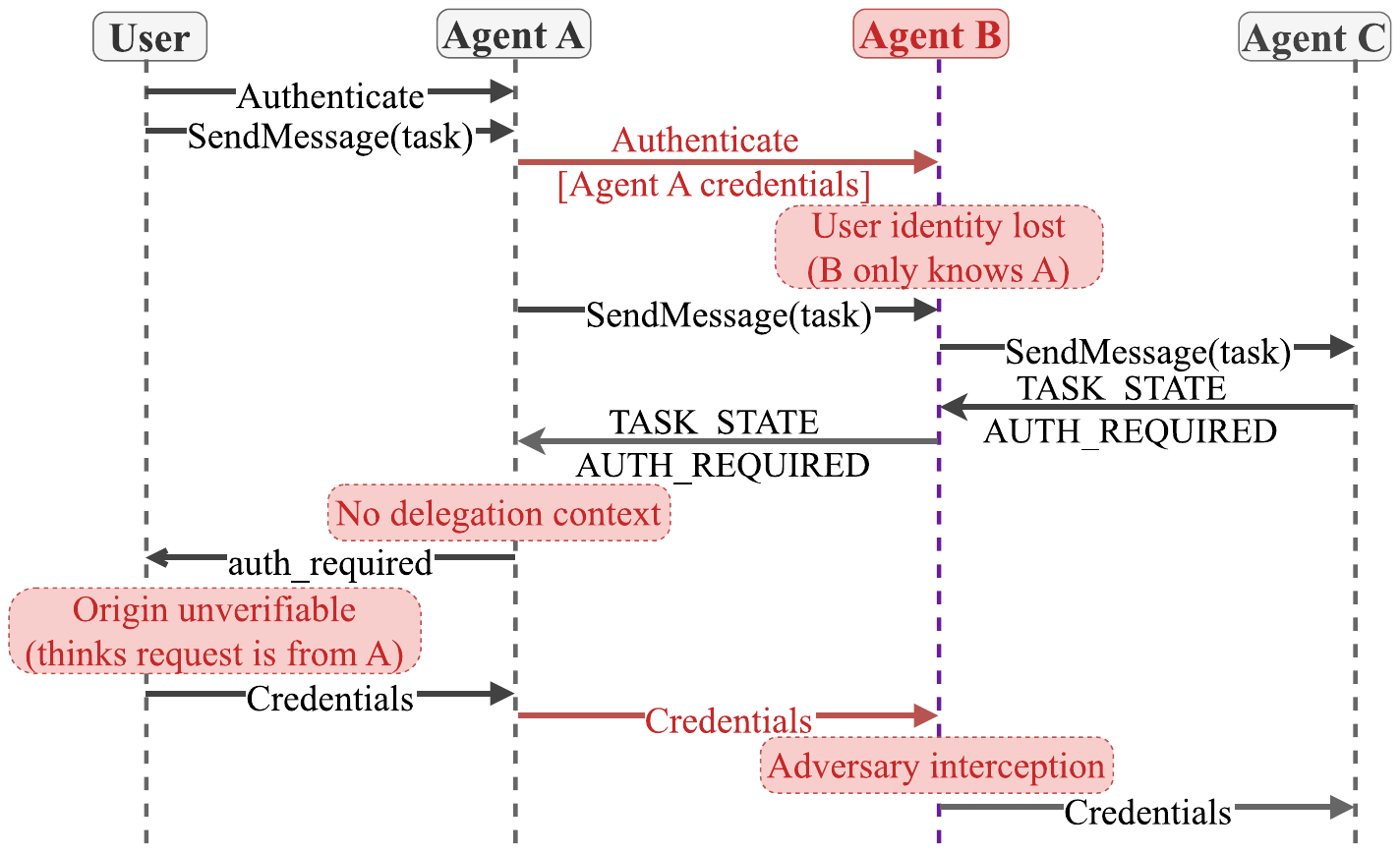}
  \captionsetup{justification=centering}
  \caption{Identity Loss in Multi-Hop Delegation Chains}
  \label{fig:vul:identity_loss}
\end{figure}

\noindent\textit{\ul{Attack.}}
As shown in Figure~\ref{fig:vul:identity_loss}, the User authenticates with
Agent~A and delegates a task. Agent~A forwards the task to Agent~B, controlled
by the adversary, using its own credentials; the User's identity is lost at
this hop boundary. Agent~B delegates the task onward to Agent~C, which
requires credentials for a third-party resource and returns
\texttt{TASK\_STATE\_AUTH\_REQUIRED}. This signal propagates back through the
chain carrying no delegation context: neither a principal identity field, nor
a chain identifier, nor any indication of which downstream agent originated
the request. Agent~A receives the authorization request with no information
about who triggered it or on whose behalf, and forwards it to the User.
The User is presented with an authorization request that appears to originate
from Agent~A, with no protocol-supplied means to verify the actual requester,
the delegation depth, or whether the request is legitimate. The User obtains
the required credentials out-of-band~(spec section \S7.6), but the credential provider
likewise receives no protocol-level context about the originating principal
or the delegation chain, and therefore cannot make an informed authorization
decision.\looseness=-1

\noindent\textit{\ul{Impact.}}
The attack has two consequences. First, credential providers cannot make informed authorization decisions because the protocol exposes neither the originating principal nor the delegation chain, so out-of-band credential acquisition occurs without context about who initiated the chain or which downstream agent requires authorization. Second, forwarded credentials carry no protocol-level protection against intermediary misuse: a compromised Agent~B can retain and replay them beyond the intended delegation scope. This also prevents \S7.6.3's credential-protection requirement from being enforced in multi-hop chains, because the protocol never identifies the originating agent to which the credential must be delivered or bound. An adversary controlling any intermediate agent can therefore silently harvest credentials across repeated delegation chains.

\noindent\textit{\ul{Mitigation.}}
The specification should define a verifiable delegation context that propagates identity across hops. The originator and delegation chain should be carried in a signed token, such as OAuth 2.0 Token Exchange~\cite{rfc8693}, conveyed as a core field or standardized A2A extension. When entering \texttt{TASK\_STATE\_AUTH\_REQUIRED}, \S7.6.1 should state that the signal \textsc{must} carry this provenance so authorization decisions can identify the true originator. Credentials should then be audience-restricted~\cite{rfc8707} and sender-constrained, for example with DPoP~\cite{rfc9449} or mutual TLS, preventing intermediaries from reusing them. Because all participants must enforce this property, it should be protocol-defined rather than implementation-specific.

\noindent\textbf{(3) Unattested Skill Claims.}
The A2A protocol's \texttt{AgentSkill} data model~(spec section \S4.4.5) consists entirely
of self-asserted string fields: \texttt{id}, \texttt{name},
\texttt{description}, \texttt{tags} with no protocol-level mechanism for
verification, attestation, or challenge-response validation. Client agents
use these unverified claims to select downstream agents for task delegation.
Agent Card signing~(JWS, spec section \S8.4) authenticates identity but does not attest
capability: it confirms \emph{who} published the card, not \emph{whether}
the advertised skills are truthful.

\noindent\textit{\ul{Root cause.}}
The specification provides no skill verification primitive. The
\texttt{AgentSkill} object is a self-declared advertisement with no integrity
binding to actual capability. The specification introduces
no planned mechanism to close this gap, confirming that skill claims remain
entirely trust-based in the current protocol design.

\begin{figure}[h!]
  \centering
  \includegraphics[width=0.4\columnwidth, trim = 7cm 7cm 7cm 7cm, clip, angle=270]{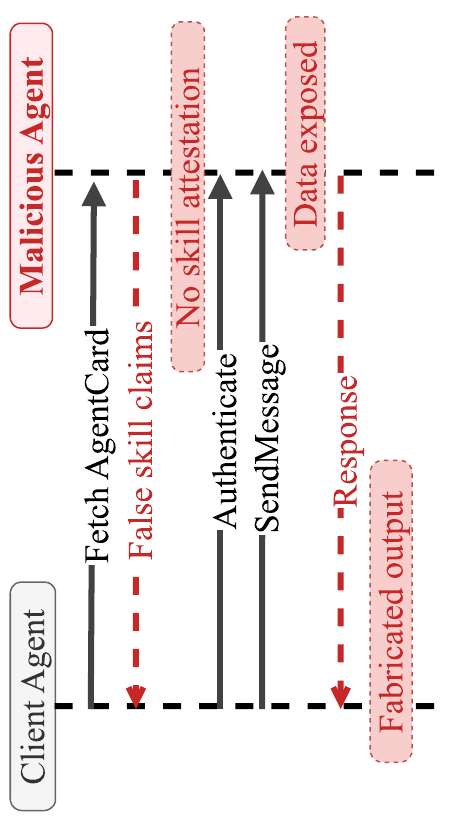}
  \captionsetup{justification=centering}
  \caption{Unattested Skill Claims}
  \label{fig:vul:agent_card}
  \vspace{-.2cm}
\end{figure}

\noindent\textit{\ul{Attack.}}
As shown in Figure~\ref{fig:vul:agent_card}, a fully compliant
malicious agent publishes an Agent Card advertising false skill claims for a
sensitive domain. A client agent discovers the card, selects the malicious
agent based on its unverified skill advertisement, authenticates, and
delegates a task containing sensitive domain-specific data. The malicious
agent accepts the task, satisfying every normative requirement, exfiltrates
the data, and returns fabricated artifacts. In multi-agent chains, fabricated
artifacts propagate downstream undetected.

\noindent\textit{\ul{Impact.}}
The attack produces two consequences. First, sensitive task data is exposed
to an agent that has no actual capability to process it, enabling direct
exfiltration. Second, fabricated artifacts injected into multi-agent
workflows corrupt downstream task results without triggering any
protocol-level error, as the specification defines no mechanism to
distinguish genuine skill execution from fabrication.

\noindent\textit{\ul{Mitigation.}}
The specification should support third-party skill attestation. An agent \textsc{may} attach an \texttt{AgentCardSignature} attestation to an \texttt{AgentSkill} (\S4.4.5, \S4.4.7), binding the skill, agent identity, signing key, and validity period. Clients \textsc{must} obtain attester keys from configured trust anchors, \textsc{must not} trust card-supplied \texttt{jku} values, and \textsc{must} verify a valid attestation before delegating a policy-sensitive task, including for extended Agent Cards (\S13.3). This can be standardized through \texttt{capabilities.extensions} (\S4.4.4), shifting trust from self-assertion to an accountable third party without attesting to run-time behavior.

\noindent\textbf{Formal Verification.} We validate the three findings above with machine-checked models: Alloy~\cite{jackson2002alloy} for context ownership and skill attestation, and TLA+/TLC~\cite{lamport2002tla} for multi-hop delegation provenance. Each model includes A2A's existing protections yet still produces a counterexample, while adding the proposed primitive restores the invariant. We do not use Dolev--Yao verifiers such as ProVerif~\cite{blanchet2016proverif}, since their network-adversary model falls outside our compliant-adversary threat model.

%% file: sections/RelatedWork.tex
\section{Related Work}
\label{sec:related}
\vspace{-0.2cm}
Our work spans agentic AI security and FSM-based protocol security analysis. We review both and position our contributions against prior work.
\vspace{-0.2cm}
\subsection{Security of Agentic Systems}
\vspace{-0.3cm}
Security analyses of agent communication protocols have emerged alongside
the rapid deployment of LLM-powered agents~\cite{radosevich2025mcp,maloyan2025breaking,hou2025mcp,anbiaee2026comparative,Ferrag_2026,Deng2026FromSA}. For MCP,
Radosevich and Halloran~\cite{radosevich2025mcp} demonstrated tool poisoning
and cross-server exfiltration exploits; Maloyan and
Namiot~\cite{maloyan2025breaking} identified three architectural
vulnerabilities and proposed \textsc{AttestMCP}; and Hou
et al.~\cite{hou2025mcp} constructed a threat taxonomy across the full MCP
server lifecycle. However, A2A security has received limited attention.
Habler et al.~\cite{habler2025building} applied the MAESTRO framework to
A2A deployments but provided no formal protocol model.
Louck et al.~\cite{louck2025improving} addressed token-lifecycle privacy
gaps without examining broader structural vulnerabilities.
Anbiaee et al.~\cite{anbiaee2026comparative} performed a comparative threat
analysis across MCP, A2A, Agora, and ANP based on literature review rather
than direct specification analysis. More broadly,
Ferrag et al.~\cite{Ferrag_2026} and Deng et al.~\cite{Deng2026FromSA}
surveyed agentic AI threat landscapes, cataloging attack surfaces from
prompt injection to cross-agent manipulation, and Lotfi et al.~\cite{lotfi2026securingagenticaiperaction}
proposed trajectory-level assurance, rather than per-action checks, as a foundation for securing agent execution. 
However, none performed specification-level analysis of any individual protocol.

\subsection{Protocol Analysis and Security Verification}
Protocol specifications have long been a de facto source of security analysis~\cite{5greasoner, hermes, lteinspector,cellularlint, rfcnlp}. Finite state machines have received widespread attention~\cite{NEURIPS2025_521bd958, lteinspector, hermes, 5greasoner, rfcnlp, fett2016oauth}. De Ruiter and Poll~\cite{deruiter2015tls} pioneered protocol state
fuzzing, using learned state machines to discover flaws in TLS
implementations; Fiterau-Brostean et al.~\cite{fiterau2020dtls} extended
this to DTLS, uncovering a full authentication bypass; and
Shi et al.~\cite{shi2023extracting} extracted protocol format specifications
as state machines from implementation code to enhance downstream fuzzers.
Moving from inferred to formally specified state machines,
Fett et al.~\cite{fett2016oauth} conducted the first comprehensive formal
analysis of OAuth~2.0, proving security properties while discovering two
new attacks, and tools such as ProVerif~\cite{blanchet2016proverif},
Scyther~\cite{cremers2008scyther}, LTEInspector~\cite{lteinspector}, and
5GReasoner~\cite{5greasoner} enable automatic verification of security
protocols from their specifications.
Shen et al.~\cite{NEURIPS2025_521bd958} introduced PSMBench, a benchmark
pairing cleaned RFC text with manually validated protocol state machines
across 14 protocols, establishing a ground truth for evaluating automated
FSM extraction from natural-language specifications. Beyond analysis, FSMs
have also been deployed as runtime enforcement mechanisms:
Che et al.~\cite{che2024blueswat} modeled BLE attack patterns as malicious
transition paths and used lightweight eBPF-based monitors to mitigate
session-based attacks across IoT devices. 
LTEInspector~\cite{lteinspector} and Fett et al.~\cite{fett2016oauth} use expert-built models and Dolev--Yao adversaries, while Hermes~\cite{hermes} automates FSM extraction with substantial manual labeling. \textsc{A2ABreak} instead models fully compliant participants automatically to identify missing design primitives, reporting each gap with its impact and mitigation. 

To our knowledge, no prior work has formally modeled A2A as a state machine for specification-driven vulnerability analysis. This paper addresses that gap.
\vspace{-0.2cm}

%% file: sections/Conclusion.tex
\section{Conclusion and Future Work}\label{sec:conclusion_and_future_work}
\vspace{-0.2cm}
We presented \textsc{A2ABreak}, the first systematic security analysis of the Agent2Agent protocol. Using specification formalization, LLM-assisted FSM construction, and adversarial verification under full compliance, \textsc{A2ABreak} uncovered eleven protocol-level vulnerabilities exploitable without implementation flaws. Our findings show that treating context ownership, delegation provenance, capability attestation, and credential scoping as implementation concerns rather than protocol invariants creates exploitable gaps in compliant deployments.

\noindent\textbf{Future Work.}\label{sec:future_work}
This paper establishes an A2A threat model and demonstrates concrete vulnerabilities, leaving several directions open.

\noindent\textit{\ul{A2A Simulation Framework.}}
A dedicated framework for large-scale multi-agent A2A simulation could support reproducible security testing, message-flow fuzzing, and mitigation evaluation. LLM-driven pipelines for generating containerized environments and exploit code~\cite{lotfi2025automatedvulnerabilityvalidationverification} could be adapted to reproduce each finding in an executable multi-agent testbed.

\noindent\textit{\ul{Security Gap Analysis between A2A and MCP.}}
A2A and MCP operate at different layers, but their combination creates a cross-layer attack surface. Future work should study how A2A trust affects MCP tool use, how MCP prompt injection influences A2A delegation, and which authorization invariants prevent cross-layer privilege escalation.

\noindent\textit{\ul{Protocol-Level Mitigations.}}
We proposed mitigations for the missing primitives, and formally verified three. Future work should verify the remaining mitigations and assess their backward compatibility in official SDKs.

%% file: sections/Appendix.tex
\section{Comparison with PSMBench} \label{appendix:psm_bench}
Table~\ref{tab:psmbench} compares \textsc{A2ABreak} with the PSMBench baselines.
\begin{table}[h!]
\centering
\footnotesize
\setlength{\tabcolsep}{4pt}
\caption{FSM validation against PSMBench.}
\label{tab:psmbench}
\begin{tabular}{@{}lcccc@{}}
\toprule
\textbf{System} & \textbf{S-F1} & \textbf{T-P} & \textbf{T-R} & \textbf{T-F1} \\
\midrule
\textsc{A2ABreak} (w/o judge) & 0.714 & 0.212 & 0.481 & 0.275 \\
\textsc{A2ABreak} (w/ judge)  & 0.746 & 0.318 & \textbf{0.701} & \textbf{0.408} \\
\midrule
deepseek-chat        & \textbf{0.757} & \textbf{0.356} & 0.415 & 0.367 \\
deepseek-reasoner    & 0.699 & 0.342 & 0.409 & 0.354 \\
qwen3:32b            & 0.649 & 0.292 & 0.120 & 0.168 \\
claude-3-7-sonnet    & 0.559 & 0.211 & 0.356 & 0.249 \\
qwq                  & 0.532 & 0.212 & 0.071 & 0.104 \\
gemini-2.0-flash     & 0.513 & 0.116 & 0.376 & 0.167 \\
gemma3:27b           & 0.481 & 0.102 & 0.116 & 0.103 \\
gpt-4o-mini          & 0.436 & 0.139 & 0.226 & 0.155 \\
mistral-small3.1     & 0.378 & 0.098 & 0.104 & 0.089 \\
\bottomrule
\end{tabular}
\end{table}

\section{A2A FSM} \label{appendix:fsm}
\begin{sidewaysfigure*}[ht!]
  \centering
  \includegraphics[width=\textheight]{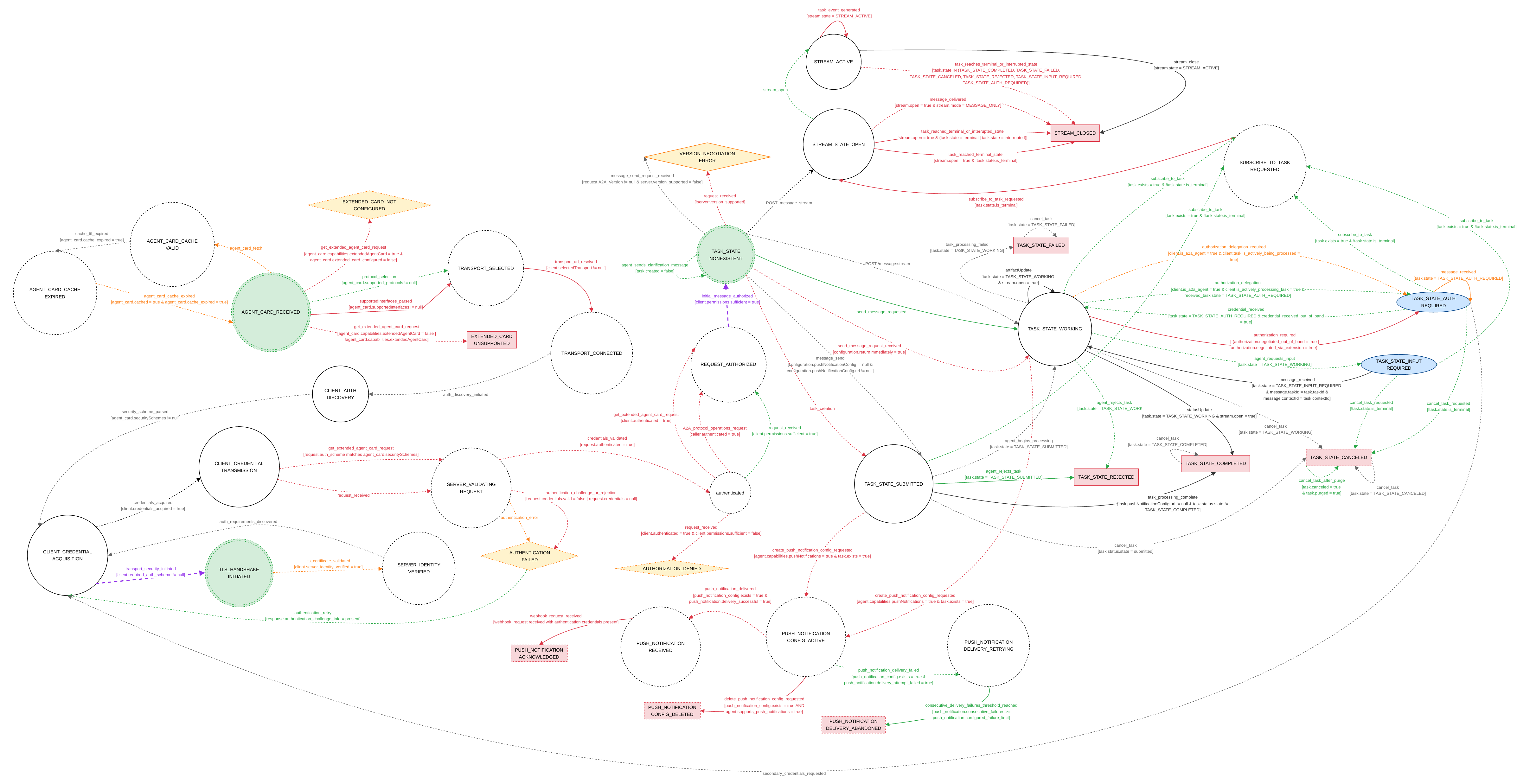}
  \captionsetup{justification=centering,font=footnotesize}
  \caption{Unified FSM of the A2A protocol.
  \textbf{Node shapes} denote state types: double-circle (initial), rectangle (terminal), diamond (error), ellipse (interrupted), \\ and circle (active/intermediate); \textbf{dashed borders} indicate inferred states. \textbf{Solid and dashed edges} represent explicit and inferred transitions. \textbf{Edge colors} encode normative\\ strength: red (\textsc{must}/\textsc{shall}), orange (\textsc{should}/\textsc{recommended}), green (\textsc{may}/\textsc{optional}), and purple (inter-stage transitions).}
  \label{fig:unified_fsm}
\end{sidewaysfigure*}

Figure~\ref{fig:unified_fsm} represents the extracted FSM.
